\documentclass[%
 reprint,
 superscriptaddress,
 amsmath,amssymb,
 aps,
 prl,
]{revtex4-2}

\usepackage{graphicx}
\usepackage{dcolumn}
\usepackage{bm}
\usepackage[hidelinks]{hyperref}
\usepackage{siunitx}

\begin{document}

\preprint{APS/123-QED}

\title{Vortex Refraction at Tilted Superconductor-Normal Metal Interfaces}

\author{Matéo F. L. Roinard-Chauvet}
\affiliation{Université Paris-Saclay, ENS Paris-Saclay, DER de Physique, 91190, Gif-sur-Yvette, France}
\affiliation{Laboratory of Nanoscale Magnetic Materials and Magnonics, Institute of Materials (IMX), School of Engineering, École Polytechnique Fédérale de Lausanne (EPFL), Lausanne, 1015, Vaud, Switzerland}
\author{Axel J. M. Deenen}
\affiliation{Laboratory of Nanoscale Magnetic Materials and Magnonics, Institute of Materials (IMX), School of Engineering, École Polytechnique Fédérale de Lausanne (EPFL), Lausanne, 1015, Vaud, Switzerland}
\author{Dirk Grundler}
\email{Dirk.Grundler@epfl.ch}
\affiliation{Laboratory of Nanoscale Magnetic Materials and Magnonics, Institute of Materials (IMX), School of Engineering, École Polytechnique Fédérale de Lausanne (EPFL), Lausanne, 1015, Vaud, Switzerland}
\affiliation{Institute of Electrical and Micro Engineering, School of Engineering, École Polytechnique Fédérale de Lausanne (EPFL), Lausanne, 1015, Vaud, Switzerland}


\date{\today}

\begin{abstract}
We derive a refraction law for superconducting vortices at superconductor/normal metal interfaces. Simulations of the proximity effect under tilted geometries confirm this law and reveal vortex trapping for low effective mass. Under transport currents, we find core displacements due to differing vortex viscosities in the superconductor and normal metal. These results clarify vortex dynamics in proximity-coupled systems and offer design principles for high-current coated superconducting devices.
\end{abstract}

\maketitle


The superconducting proximity effect refers to the leaking of Cooper pairs from a superconducting material (S) into a non-superconducting (normal) material by placing the two in direct contact \cite{de_gennes_superconductivity_1999}. This phenomenon has enabled rich and unconventional physics in a variety of systems, including normal metals \cite{clarke_proximity_1968}, ferromagnets \cite{buzdin_proximity_2005, linder_superconducting_2015}, and topological insulators \cite{fu_superconducting_2008}. In normal metals (N), proximity-induced correlations can be long-ranged, persisting over several hundred nanometers.

Of particular interest is the emergence of topological defects of the superconducting order parameter $\psi$, such as Abrikosov vortices, in normal metals with proximity-induced superconductivity \cite{peroz_vortex_2002, cuevas_magnetic_2007, amundsen_field-free_2018}. Until recently, the evolution of a vortex transitioning from a superconductor into a normal metal has not been well understood. Stolyarov et al. \cite{stolyarov_expansion_2018} demonstrated experimentally, supported by quasiclassical calculations, that when a superconductor in the mixed state is brought into contact with a diffusive normal metal under a perpendicular magnetic field, the vortex core penetrates the normal region and expands via the proximity effect, a result confirmed by Panghotra et al. \cite{panghotra_exploring_2019}. More recently, Vodolazov \cite{vodolazov_squeezed_2024}, using time-dependent Ginzburg-Landau simulations, reported the formation of nascent vortices at the S/N interface under an in-plane magnetic field. Earlier theoretical and numerical works also explored vortex interactions with point, tilted line, and plane defects in high-temperature superconductors \cite{blatter_vortices_1994, kwok_vortices_2016}. However, a general understanding of vortex behavior in S/N bilayers for arbitrary interface orientations with respect to the applied field remains lacking. Additionally, the behavior of proximity induced vortices under transport currents is not understood.

In this letter, we investigate how vortices interact with superconductor-normal metal interfaces at arbitrary orientations. We show that vortices crossing a tilted interface undergo a bending of their core, depending on the discontinuity in the effective Cooper pair mass. In analogy with the terminology in other fields, such as fluid dynamics \cite{lugt_oblique_1989} and electromagnetism \cite{lorrain_electromagnetic_1996}, we refer to this phenomenon as refraction. Using analytical arguments, we derive the corresponding vortex refraction law, and, through time-dependent Ginzburg-Landau simulations, we demonstrate that this effect, combined with interface energy barriers, leads to an apparent vortex core displacement in static conditions. We show that applied transport currents modulate the degree of vortex bending. Moreover, we find that tilting the interface introduces directional asymmetry in vortex motion, which manifests as changes in voltage oscillation frequency.

Proximity-induced superconductivity in the metal is described using a modified Ginzburg-Landau framework following \cite{chapman_ginzburglandau_1995}. This approach corresponds to modeling the normal metal as a superconductor above its critical temperature. In the following, subscripts S and N designate parameter values taken in the superconductor and normal metal, respectively. The evolution of the order parameter $\psi$ is described using the time-dependent Ginzburg-Landau (TDGL) equation as given by \begin{multline}
\frac{1}{D_{\mathrm{S/N}}}\left(\frac{\partial\psi}{\partial t}+\mathrm{i}\kappa\phi\psi\right)\\+\left(\frac{\mathrm{i}}{\kappa}\nabla+\mathbf{A}\right)^2\psi + (a_{\mathrm{S/N}} + b_{\mathrm{S/N}}|\psi|^2)\psi=0.
\label{eq:TDGL}\end{multline}
 Here, $\phi$ is the scalar potential, $\mathbf{A}$ is the vector potential, and $D_\mathrm{S/N}$ denotes the electronic diffusion constant normalized to $D_\mathrm{S} = 1$. The parameters $a_\mathrm{S/N}$ and $b_\mathrm{S/N}$ are normalized such that $a_\mathrm{S}=-1$, $b_\mathrm{S}=1$, $a_\mathrm{N}>0$, and $b_\mathrm{N}=0$. The parameter $a_\mathrm{N}$ is taken as phenomenological and adjusted accordingly. We denote $\kappa = \lambda/\xi$ the Ginzburg-Landau parameter with $\lambda$ the London penetration depth and $\xi$ the coherence length. We use the following units: $[r] = \lambda$, $[t] = \tau = \xi^2/D_\mathrm{S}$, $[\mathbf{A}] = \Phi_0/(2\pi\xi)$, and $[\phi] = V_0 = \Phi_0D_\mathrm{S}\lambda/(2\pi\xi^3)$, with $\Phi_0$ the magnetic flux quantum. The order parameter $\psi$ is scaled to its bulk value at zero magnetic field.
 Our simulations showed that for the geometrical parameters considered here, the screening field did not significantly alter the vortex shape in either the superconductor or the metal, as shown in the Supplementary Material~\cite{SuppMat}. In the following, we therefore assume a constant vector potential $\mathbf{A}$ and only solve for Eq.~(\ref{eq:TDGL}) coupled with the Poisson equation for the scalar potential: \begin{equation}
    \Delta\phi = -\frac{\mathrm{i}}{2m_\mathrm{S/N}\kappa\sigma_\mathrm{S/N}}\nabla(\psi^*\nabla\psi-\psi\nabla\psi^*),
\end{equation} where the Coulomb gauge $\nabla\cdot\mathbf{A}=0$ is assumed, with $\sigma_\mathrm{S/N}$ the normal-state conductivity, and $m_\mathrm{S/N}$ the effective mass of Cooper pairs, taking $m_\mathrm{S} = 1$ by normalization. Boundary conditions at the interface with the insulator surrounding the system are: $\nabla\psi\cdot\mathbf{n} = 0$, $\sigma_\mathrm{S/N}\nabla\phi\cdot\mathbf{n} = -\mathbf{J}\cdot\mathbf{n}$, $\mathbf{A}\cdot\mathbf{n} = 0$ and $(\nabla\times\mathbf{A})\times\mathbf{n}=\mathbf{H}\times\mathbf{n}$ with $\mathbf{n}$ the normal to the interface, $\mathbf{J}$ the applied transport current and $\mathbf{H}$ the applied magnetic field. The condition $\psi = 0$ is imposed on the surfaces where current is injected.
Following \cite{chapman_ginzburglandau_1995}, at the interface between the superconductor and the normal metal, boundary conditions are $[\psi]=0$, $[\mathbf{A}]=\mathbf{0}$, $[\phi]=0$, $[(\nabla\times\mathbf{A})\times\mathbf{n}/\mu_\mathrm{S/N}]=\mathbf{0}$, $[\sigma_\mathrm{S/N}(\partial_t\mathbf{A}+\nabla\phi)\cdot\mathbf{n}]=0$ and \begin{equation}[\mathbf{n}\cdot\frac{1}{m_\mathrm{S/N}}(\mathrm{i}\nabla/\kappa+\mathbf{A})\psi] = 0\label{eq:dpsi_jump},\end{equation} where $\mu_\mathrm{S/N}$ is the magnetic permeability, and $[\cdot]$ denotes the jump of the enclosed quantity across the interface.

\begin{figure}[t]
\includegraphics{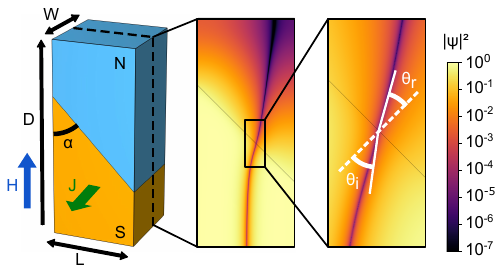}
\caption{An oblique view of the system (left): a superconducting (S) - normal metal (N) bilayer with a tilted interface described by the angle $\alpha$ subject to an applied dc current density $\mathbf{J}$ parallel to the boundary and a vertical magnetic field $\mathbf{H}$. A cross-sectional view of the system reveals a vortex refracted at the interface (center) which permits to define incident and refraction angles $\theta_\mathrm{i}$ and $\theta_\mathrm{r}$, respectively (right). The two panels on the right show the color-coded order parameter $|\psi|^2$ (see legend) for which bright (black) encodes the superconducting state (normal state).}
\label{Fig_SN_1}
\end{figure}

To investigate the orientation and shape of vortices at a tilted boundary, we performed numerical simulations of a S/N bilayer structure (Fig.~\ref{Fig_SN_1}) using the finite element method \cite{gao_finite_2023} implemented in COMSOL \cite{noauthor_comsol_nodate, ADeenen2025, oripov_time-dependent_2020}. The system dimensions are $h = 3.5\lambda$, $L = W = 1.5\lambda$, with the interface tilted by an angle $\alpha$ relative to the vertical. The superconductor considered is Nb in the dirty limit, a widely used type II material in proximity effect studies, with a coherence length \( \xi(0) = \SI{9}{\nano\meter} \), and a penetration depth \( \lambda(0) = \SI{90}{\nano\meter} \), corresponding to a Ginzburg-Landau parameter $\kappa = 10$ \cite{stolyarov_expansion_2018, krasnov_anomalous_1992}. We used \( \sigma_\mathrm{S} = \SI{15.5}{(\micro\ohm\meter)^{-1}} \), \( D_\mathrm{S} = \SI{1.3e-3}{\meter\squared\per\second} \) for the superconductor, and \( \sigma_\mathrm{N} = \SI{27}{(\micro\ohm\meter)^{-1}} \),  \( D_\mathrm{N} = \SI{0.94e-3}{\meter\squared\per\second} \) for the normal metal (Cu) \cite{stolyarov_expansion_2018}. From microscopic theory \cite{giorgi_jadallah_2006}, the effective mass parameter of the Cooper pairs is given by $m_\mathrm{N}/m_\mathrm{S} = \sigma_\mathrm{S}/\sigma_\mathrm{N} = 0.65$. To understand the effect of this parameter on the anticipated vortex bending in further detail, we also performed simulations with $m_\mathrm{N} = 10$ and $m_\mathrm{N} = 0.01$. The coefficient $a_\mathrm{N}$ was set to $0.003$ to reproduce the shape of proximity induced vortices in the normal metal consistent with \cite{stolyarov_expansion_2018}. In all the following, a magnetic field $H = 1$ in the given units is applied vertically.

Figure~\ref{Fig_SN_1} shows a vortex bending at the S/N interface, characterized by an incident (i) angle $\theta_\mathrm{i}$ and a refracted (r) angle $\theta_\mathrm{r}$. To understand the behavior observed in the simulations, we develop an analytical model describing the refraction-like bending of a vortex line crossing an interface tilted by an angle $\alpha$ with respect to the vertical (Fig.~\ref{Fig_SN_1}). Near the vortex core, the order parameter modulus increases linearly with distance to the core $r$ \cite{tinkham_introduction_1996}, allowing the approximation $|\psi| = \gamma_\mathrm{S/N} r$, where $\gamma_\mathrm{S}$ and $\gamma_\mathrm{N}$ are constants. From Eq.~\ref{eq:dpsi_jump}, at the vicinity of the vortex core where $\psi$ tends linearly to $0$, we obtain $\mathbf{n}\cdot\nabla\psi_\mathrm{S}/m_\mathrm{S}=\mathbf{n}\cdot\nabla\psi_\mathrm{N}/m_\mathrm{N}$, and therefore, \begin{equation}\frac{1}{m_\mathrm{S}}|\nabla\psi_\mathrm{S}|\sin(\theta_\mathrm{i}) = \frac{1}{m_\mathrm{N}}|\nabla\psi_\mathrm{N}|\sin(\theta_\mathrm{r}),\label{eq:NablaPsiNormal}\end{equation} where the angles $\theta_\mathrm{i}$ and $\theta_\mathrm{r}$ are defined in Fig.~\ref{Fig_SN_1}. Continuity of the order parameter across the interface gives $\gamma_\mathrm{S}\cos(\theta_\mathrm{i}) = \gamma_\mathrm{N}\cos(\theta_\mathrm{r})$, from which we deduce \begin{equation}|\nabla\psi_\mathrm{S}|\cos(\theta_\mathrm{i}) = |\nabla\psi_\mathrm{N}|\cos(\theta_\mathrm{r}).\label{eq:NablaPsiTangent}\end{equation} Dividing equation \ref{eq:NablaPsiNormal} by equation \ref{eq:NablaPsiTangent}, we get the refraction law 

\begin{equation}\frac{1}{m_\mathrm{S}}\tan(\theta_\mathrm{i}) = \frac{1}{m_\mathrm{N}}\tan(\theta_\mathrm{r}).\label{eq:refractionLaw}\end{equation}  

\begin{figure}[t]
\includegraphics{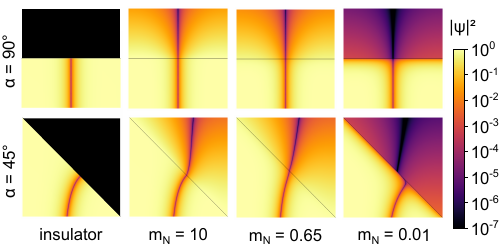}
\caption{Cross-sectional view of the crossing of a superconductor/normal metal interface by a vortex for various effective Cooper pair mass in the metal $m_\mathrm{N}$. The metal is on top of the superconductor.}
\label{Fig_StaticSimul_2}
\end{figure}

Figure~\ref{Fig_StaticSimul_2} shows cross-sectional views of the vortex line at the S/N interface for various angles $\alpha$ and effective masses $m_\mathrm{N}$ in the static case with $J = 0$. The top row corresponds to $\alpha = 90^\circ$, where the applied magnetic field is orthogonal to the interface. In that case, the vortex crosses the boundary without tilting for all masses. The bottom row shows the distribution of the order parameter for $\alpha = 45^\circ$. Here, the vortex is deviated when crossing the interface, with a deviation angle $\theta_\mathrm{i}-\theta_\mathrm{r}$ depending on the mass parameter $m_\mathrm{N}$ of the metal. The leftmost column shows the results obtained for an insulator in place of the metal. This corresponds to the limit $m_\mathrm{N}\to\infty$ and $a_\mathrm{S}\to\infty$ with $\sqrt{a_\mathrm{S}}/m_\mathrm{S}\to0$ \cite{chapman_ginzburglandau_1995}. Boundary condition $\mathbf{n}\cdot\nabla\psi=0$ with the insulator imposes a vortex core perpendicular to the boundary, as observed in Fig.~\ref{Fig_StaticSimul_2}.

\begin{figure}[b]
\includegraphics{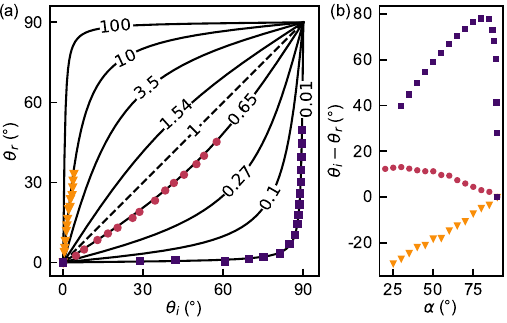}
\caption{(a) Refraction angle $\theta_\mathrm{r}$ as a function of incident angle $\theta_\mathrm{i}$ for various effective Cooper pair mass $m_\mathrm{N}$ in the metal at selected values for the effective mass. Solid lines correspond to analytical refraction law solutions (Eq. (\ref{eq:refractionLaw})), with $m_\mathrm{N}$ indicated on top, and symbols are results extracted from simulations. (b) Deviation angle $\theta_\mathrm{i}-\theta_\mathrm{r}$ as a function of interface inclination $\alpha$. The color coding indicates different values of $m_\mathrm{N}$ consistent with (a).}
\label{Fig_ComparisonSimuModel_3}
\end{figure}

Figure~\ref{Fig_ComparisonSimuModel_3}a compares refraction angles $\theta_\mathrm{r}$ extracted from simulations (symbols) and analytical predictions based on Eq.~(\ref{eq:refractionLaw}) (lines) for various $m_\mathrm{N}$. In contrast to refraction of electromagnetic waves in the optical regime, there is no critical angle beyond which the refraction of vortices ceases. The deviation angle $\theta_\mathrm{i} - \theta_\mathrm{r}$ vanishes in both normal and grazing incidence. However, not all possible incident angles could be reached in the simulation due to the vortex leaving the system when $\alpha$ became too small. Reaching more incident angles should be possible in a larger system or under stronger magnetic fields. Still, the excellent quantitative agreement between numerical and analytical models shown in Fig.~\ref{Fig_ComparisonSimuModel_3}a motivates broader FEM exploration beyond analytical reach, particularly in dynamic regimes.

Decreasing the angle $\alpha < 90^\circ$ increases both the incident and refraction angles, as the vortex tends to align with the applied magnetic field. However, repulsion from the SN boundary in the superconductor favors perpendicular vortex alignment, reducing $\theta_\mathrm{i}$. This effect is stronger for a larger jump in effective Cooper pair mass, where the vortex requires longer distances to reorient along the magnetic field. The refraction angle is less affected due to weaker boundary interaction arising from reduced Cooper pair density. Figure~\ref{Fig_ComparisonSimuModel_3}b shows the relationship between the angle $\alpha$ and the deviation angle $\theta_\mathrm{i} - \theta_\mathrm{r}$ for different mass parameters. Considering Eq.~(\ref{eq:refractionLaw}), it is clear that the deviation angle is negative when $m_\mathrm{N}$ is larger than $m_\mathrm{S}$, and positive in the opposite case. Having $\theta_\mathrm{i}-\theta_\mathrm{r}=0$ corresponds to a straight vortex.

In the insulating limit $m_\mathrm{N}\to\infty$, Eq.~(\ref{eq:refractionLaw}) yields normal incidence ($\theta_\mathrm{i} = 0$), consistent with the insulator boundary conditions. In the opposite limit $m_\mathrm{N} \ll m_\mathrm{S}$, corresponding to highly conductive metals, Eq.~(\ref{eq:refractionLaw}) predicts $\theta_\mathrm{i} = 90^\circ$ for all $\theta_\mathrm{r}\neq 0$. Correspondingly, the simulated Cooper pair density $|\psi|^2$ in Fig.~\ref{Fig_StaticSimul_2} for $m_\mathrm{N} = 0.01$ and $\alpha = 45^\circ$ shows that the vortex core tends to remain at the interface before entering the superconducting region, resulting in an apparent displacement between the vortex core in the metal and in the superconductor. The feature resembles the Goos-Hänchen effect in light transmission \cite{wang_large_2005}. In the present system, the large difference in Cooper pair density between the normal and the superconducting region creates an energy barrier similar to the Bean-Livingston barrier \cite{bean_surface_1964}, which pins a segment of the vortex along the interface \cite{vodolazov_squeezed_2024}. The vortex eventually enters the superconductor when its line energy, proportional to its length \cite{tinkham_introduction_1996}, outweighs the barrier. As $\alpha<90^\circ$ decreases, the reduced line energy cost favors longer interfacial segments, thereby enhancing this displacement. This effect is similar to vortex trapping by twin planes \cite{blatter_vortices_1994}.

\begin{figure*}[t]
\includegraphics{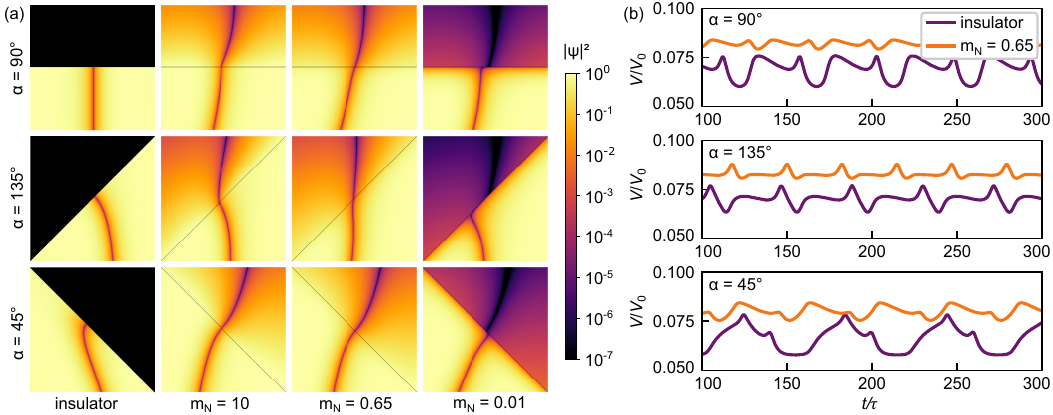}
\caption{Close views of the crossing of a superconductor/normal metal interface by a vortex under an applied transport current for various effective Cooper pair masses $m_\mathrm{N}$ in the metal (a). Vortices are moving to the right, inducing voltage oscillations (b).}
\label{Fig_TransportCurrent_4}
\end{figure*}

In the following, we discuss the influence of a constant transport current $J=0.27$, applied as shown in Fig.~\ref{Fig_SN_1}, on the position of the vortex near the interface. The transport current drives vortex motion to the right by exerting a Lorentz force on it. Figure~\ref{Fig_TransportCurrent_4}a shows cross-sections of moving vortices for various angles $\alpha$ and mass $m_\mathrm{N}$.

Remarkably, even for $\alpha = 90^\circ$, where no static tilt is observed in Fig.~\ref{Fig_StaticSimul_2}, a dynamic tilt emerges in the presence of the normal metal. This tilt arises from two key mechanisms in the metal: reduced vortex viscosity and preferential vortex nucleation. Vortex viscosity stems from energy dissipation associated with vortex motion. Here, two dissipation phenomena are taken into account \cite{schmid_time_1966}: normal current and induced electric field dissipation, and the relaxation of the order parameter. The dissipated power density associated with the latter in the given units is $W_\mathrm{T}=(1/D_\mathrm{S/N})|(\partial_t+\mathrm{i}\kappa\phi)\psi|^2$. Here, $D_\mathrm{N} > D_\mathrm{S}$, resulting in lower dissipation in the metal and hence, reduced viscosity. The power density dissipated via normal currents is given by $W_\mathrm{BS}=\sigma_\mathrm{S/N}(\partial_t\mathbf{A}+\nabla\phi)^2$, with $\sigma_\mathrm{N} > \sigma_\mathrm{S}$ here which increases dissipation and thus vortex viscosity in the metal. For the values of $D_\mathrm{S/N}$ and $\sigma_\mathrm{S/N}$ considered here, the reduced dissipation from order parameter relaxation dominates, leading overall to decreased viscosity in the metal and enhanced vortex motion there. Consequently, the vortex line bends toward the direction of motion in the normal metal. See the Supplementary Material \cite{SuppMat} for more details on those two mechanisms in our system.
Additionally, the lower Cooper pair density in the metal reduces the nucleation barrier, favoring vortex entry from the metal, further contributing to the observed tilt.

For low values of the mass parameter (e.g., $m_\mathrm{N}=0.01$, Fig. \ref{Fig_TransportCurrent_4}a), the previously discussed displacement of the vortex core between normal and superconducting regions exists, even when $\alpha = 90 ^\circ$. This results from the same mechanism as described in the static regime (Fig. \ref{Fig_StaticSimul_2}). Note, however, that this displacement in the static regime could only be realized for $\alpha\neq90$. The combined effect of applied current and varying angle $\alpha$ further modulates this displacement: $\alpha > 90^\circ$ enhances it, while $\alpha < 90^\circ$ reduces it.
The angle $\alpha$ also affects the vortex speed, leading to variations in the frequency of voltage oscillations, as shown in Fig.~\ref{Fig_TransportCurrent_4}b. Here, voltage is defined as $V = (1/S_\mathrm{A})\int_{S_\mathrm{A}}\phi\,dS - (1/S_\mathrm{B})\int_{S_\mathrm{B}}\phi\,dS$ with $S_\mathrm{A}$ and $S_\mathrm{B}$ the areas of the two surfaces where current is injected. Two competing mechanisms are at play. First, for $\alpha < 90^\circ$, the vortex length inside the superconductor decreases during its motion, reducing its line energy and creating a geometric force that favors motion, as described in \cite{moll_geometrical_2025}. Second, the taller edge where the vortex nucleates constitutes a higher energy barrier preventing vortex entry. In our small system, the edge effect dominates, resulting in reduced vortex mobility as $\alpha$ increases. Figure~\ref{Fig_TransportCurrent_4}b also shows an increased vortex mobility with the metal compared to the insulator.

In summary, we have demonstrated that a vortex penetrating a normal metal via proximity effect exhibits refraction, governed by a quantitative law derived analytically and validated through three-dimensional time-dependent Ginzburg-Landau simulations. In the static regime, vortex refraction, in combination with interfacial energy barriers and vortex line tension, results in an apparent displacement of the vortex core across the S/N boundary. Under transport current, modulation of the vortex tilting emerges from the interplay between altered vortex viscosity and preferential nucleation driven by the proximity effect. The vortex refraction law presented in this work shares deep analogies with other domains of physics, such as heat transfer \cite{tan_tangent_1990}, fluid dynamics \cite{lugt_oblique_1989, sieniutycz_variational_2007}, and electro- and magnetostatics \cite{lorrain_electromagnetic_1996}, offering a unifying perspective across physical systems. Our results provide important insights into the proximity effect of type-II superconductors under applied fields and currents. They pave the way towards an understanding of proximity effects in nonplanar geometries, which we expect to gain importance in future 3D superconducting nanostructures with engineered curvatures \cite{makarov_new_2022}.

\begin{acknowledgments}
The simulations have been performed using the facilities of the Scientific IT and Application Support Center of EPFL. We acknowledge support by the SNSF via project 10000845.
\end{acknowledgments}



\bibliographystyle{apsrev4-2}
\bibliography{biblio}

\end{document}